\documentclass{article}
\usepackage{psfig}
\begin{document}
\parskip1em\parindent0pt
\Large

\centerline{\bf CONSTRAINTS ON SIDM WITH FLAVOR MIXING}
\centerline{\it Mikhail Medvedev (CITA)}
\vskip0.5cm

The self-interacting dark matter (SIDM) model with flavor mixing 
 ({\tt astro-ph/0010616}) was proposed to resolve problems  of the 
CDM model on small scales by keeping attractive features of both 
SIDM and annihilating dark matter, and simultaneously avoid their
drawbacks. A dark particle produced in a flavor eigenstate will 
separate into two mass eigenstates because they propagate with
different velocities and, in a gravitational filed, along different 
geodesics, see Fig. 1. Thus, in the flavor-mixed SIDM, dark halos are 
made of heavy eigenstates, whereas light eigenstates may leave the 
halo. Collisions (elastic scattering) of mass states results in 
eigenstate conversion, see Fig. 2, which leads to the gradual 
decrease of the halo mass in high-density cusps. On the other hand, 
in the early Universe, one may expect a problem of over-production 
of light (hot) particles over heavy (cold) when the temperature of 
dark matter falls below the mass of the heavy component. We show 
how this problem is avoided.
\vskip1cm
\mbox{\psfig{file=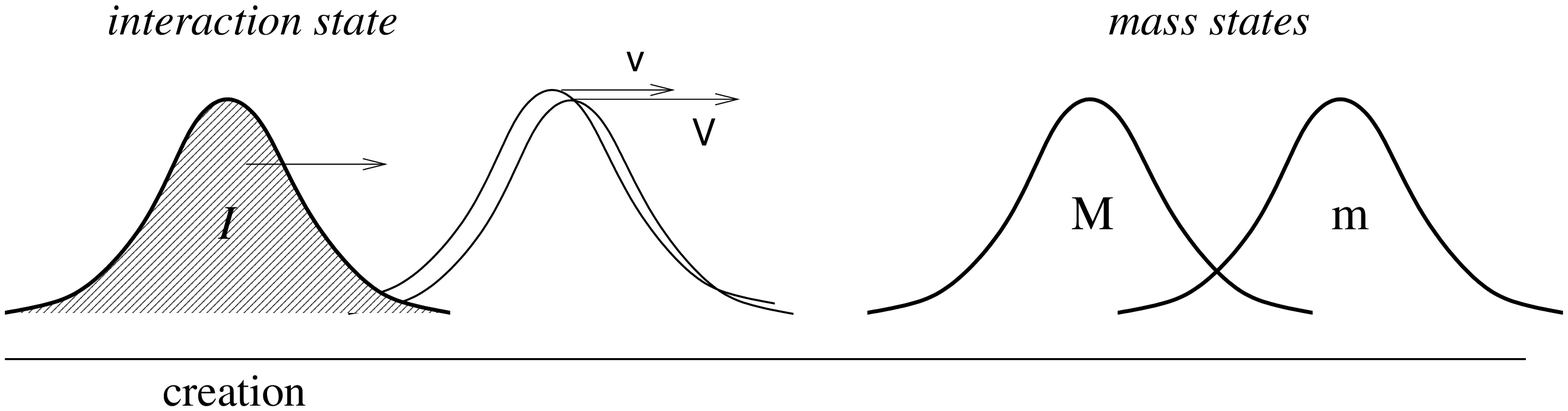,width=4.5in}~~~~~Fig.~1} \\[1em]
\mbox{\psfig{file=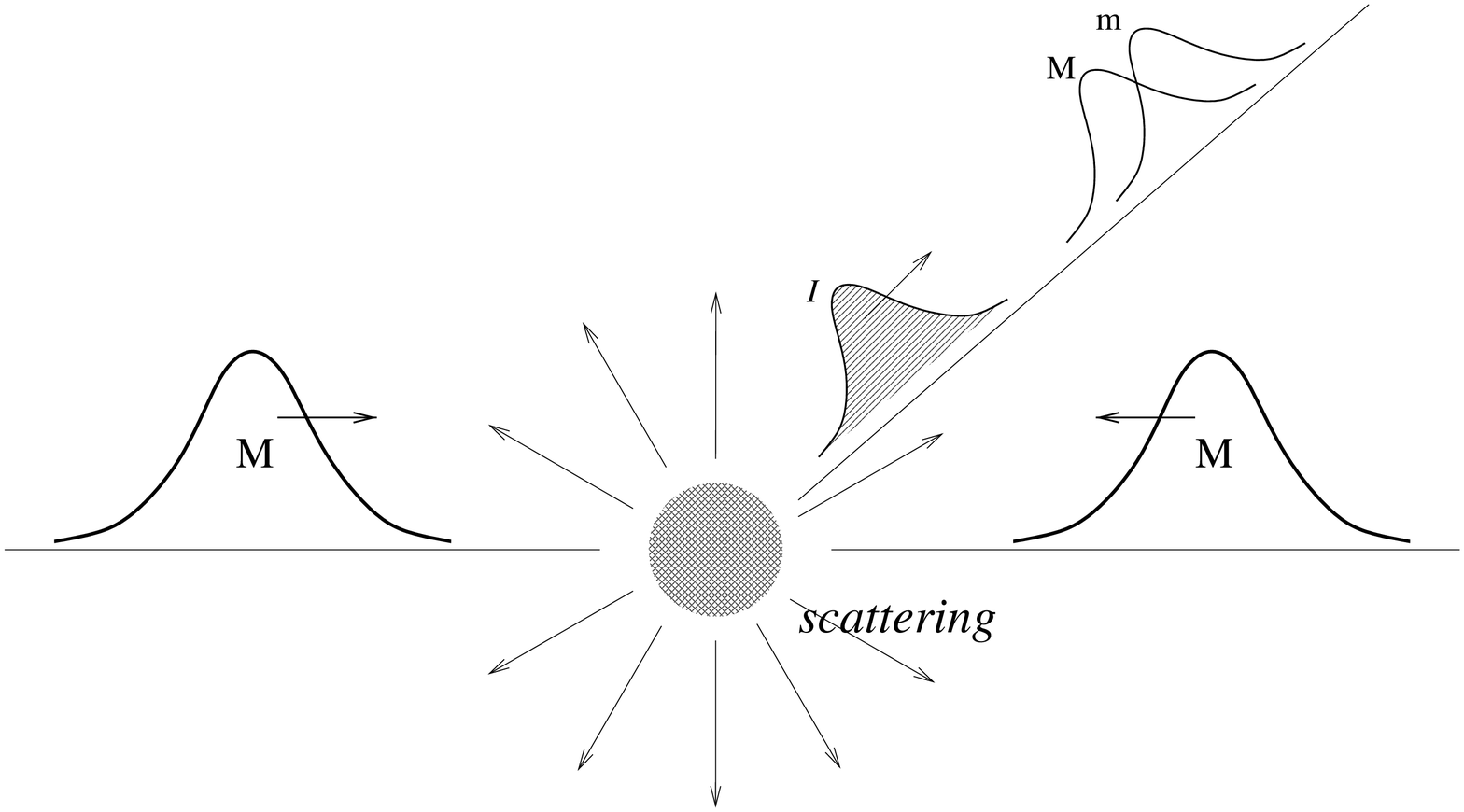,width=4.5in}~~~~~Fig.~2}

\newpage
		\centerline{\bf Constraints on the Model}		
{\it
$\bullet$ There are constraints imposed on the model by 
processes in the early Universe. In particular, when 
temperature drops below the mass difference, the number
of heavy particles is suppressed exponentially. Dark matter
then will consist predominantly of light particles which
ruins the model.
}

\vskip1em
{\bf I. Two ways to avoid over-production of light mass states}

1. If spreading of a wave packet (because of quantum uncertainty 
in the momentum of a particle at the moments of creation and 
detection) is faster than the separation of mass states due to 
the velocity difference, then the two mass states, essentially, 
never separate. In this case, DM particles created in an interaction 
state always remain in this interaction state. Such DM particles in 
the early Universe behave as a single-species collisional dark matter,
i.e., the standard SIDM. 			\hfill\break
The mass states will be separated later in the gravitational
potential of assembling dark halos, because these states propagate
along different geodesics. At some point, light states will escape
from the halo and heavy states remain bound. At even later times,
the density in cores becomes so high that collisions and conversions
become significant. Conversions will decrease the central density.

2. In the opposite case, when particles do separate, it seems the 
only way to avoid over-production is what you've suggested:
the degenerate case: $m_h\approx m_l$. Assuming that the DM-matter 
cross-section $\sigma_m$ is much smaller than the DM self-interaction 
cross-section $\sigma_{si}\gg\sigma_m$, we have that after the moment
when DM particles (both flavors) freezes-out 
(when $n_f\langle\sigma_m v_f\rangle\sim H_f$),
these particles remain highly collisional and conversions go
both ways, i.e., $m_h\to m_l$ and $m_l\to m_h$. We then need to 
require that $m_h-m_l<T_c$ at the moment when conversions halt, 
i.e., $n_c\langle\sigma_{si} v_c\rangle\sim H_c$.

Let's consider these two cases separately.

\vskip1em\vfill\break
{\bf II. Wave packet spreading}

Assuming a Gaussian wave packet at $t=0$ having the velocity
$v_0<c$ and width $a_0$, 
$$
\Psi(0)=A\exp\left(-{x^2\over 2a_0^2}+{i m v_0 x\over \hbar}\right),
$$
one can show that, at some time $t$,
$$
\left|\Psi(t)\right|^2
={|A|^2 \over \sqrt{1+{\hbar^2 t^2 \over m^2 a_0^4}}}
\exp\left[-{(x-v_0 t)^2 \over a_0^2
\left(1+{\hbar^2 t^2 \over m^2 a_0^4}\right)}\right].
$$
That is, the wave packet moves with a constant velocity $v_0$
and spreads in space as
$$
\Delta = \sqrt{\langle(x-\langle x\rangle)^2\rangle}
={a_0 \over \sqrt 2}\left(1+{\hbar^2 t^2 \over m^2 a_0^4}\right)^{1/2}
\approx {\hbar t \over \sqrt{2} m a_0} 
= {\hbar t \over 2 m \Delta_0} ~~{\rm as}~~ t\to\infty.
\eqno{(1)}
$$
This is the main equation of this section.

The wave-packet width, $\Delta_0$, is, in general, determined by 
the processes of production and detection, which are simply 
scattering events in our case. For low-energy elastic scattering,
the scattering cross-section is isotropic and does not depend on 
energy, $\sigma=4\pi l^2$. Here $l$ is the scattering length 
determining the wave packet width, $\Delta_0\sim l$.

For the cross section we use the SIDM parameterization
$$
\sigma_{si}=10^{-24}\, \sigma_{si,-24}\, m_{\rm GeV} ~{\rm cm^2},
\eqno{(2)}
$$
where $m_{\rm GeV}$ is the mass of DM particle in units of GeV.

\vskip1em\vfill\break
{\bf III. When do mass states not separate in the early Universe?}

We now consider the first scenario: the mass eigenstates
do not separate from each other in the early Universe
(i.e., in the absence of gravitational forces) but do 
separate in dark halos. There are two cases.

\vskip1em
{\it 1. Non-relativistic degenerate case}

Here we consider the case when $m_h\approx m_l\approx m$,~
$\delta m = m_h - m_l \ll m$ .
First, the widths of the wave packets are nearly the same,
$$
\Delta_h\approx\Delta_l
\approx {\hbar t \over m}{\sqrt{4\pi \over \sigma_{si}}}.
$$
Second, from the energy conservation in conversions and 
assuming $v_h\ll v_l$, we have 
$$
v_l \approx c \sqrt{2\,(\delta m)/m}.
$$
Non-separation of wave packets means that they remain overlapped
for all times, i.e.,  the distance between them is smaller than or 
comparable to the packets' width
$$
\delta x=|v_l-v_h|t \approx v_lt \approx ct\sqrt{2{\delta m\over m}}
\le \Delta \approx {\hbar t \over 2 m}{\sqrt{4\pi \over \sigma_{si}}}.
$$
This yields the constraint on masses
$$
{\delta m \over m}\le {\pi \hbar^2 \over 2 m^2 c^2 \sigma_{si}}
~~~{\rm or}~~~ 
{\delta m \over m}\le 6.1\times10^{-4}\, 
m_{\rm GeV}^{-3}\, \sigma_{si,-24}^{-1}.
\eqno{(3)}
$$
If this constraint is satisfied, the mass eigenstates do not 
separate from each other (in the absence of gravitational field) 
and their wave packets remain overlapped all the time.

For the state conversions to occur in dark halos, the mass states
must separate (segregate) in the gravitational potential,
i.e., they must move along significantly different geodesics.
This implies that the light state must be nearly unbound
whereas the heavy state be bound. Thus the velocity $v_l$
must be comparable to or greater than the typical escape 
velocity. We assume it to be $v_l \ge 10^{-3}\, c\, v_{-3}$.
This gives the lower bound on the mass difference
$$
{\delta m \over m} \ge 5\times10^{-7}\, v_{-3}^2.
\eqno{(4)}
$$
This is a complementary condition to (3). These two constraints 
can be combined together to yield the upper bound on the mass
of a DM particle in this scenario:
$$
m \le 11 \left( \sigma_{si,-24} v_{-3}^2 \right)^{-1/3}~~{\rm GeV}.
\eqno{(5)}
$$

\vskip1em
{\it 2. Non-degenerate case}

In this case $m_h \gg m_l$, i.e., $m_l$ is essentially zero and
$m_h = m$. The heavy particle is non-relativistic, therefore
$v_h \ll v_l = c$. Thus the states separate with the velocity
$\sim c$. The non-relativistic wave packet also spreads according 
to equation (1), while the relativistic packet quickly spreads
in the directions perpendicular to the direction of motion but
the spreading in the direction of motion is much smaller due to 
Lorentz contraction effects. To avoid separation of states we 
again require that
$$
\delta x = ct \le \Delta 
\approx {\hbar t \over 2 m}{\sqrt{4\pi \over \sigma_{si}}}.
$$
This puts the upper bound on mass
$$
m \le 0.10\,\sigma_{si,-24}^{-1/3}~~{\rm GeV}.
\eqno{(6)}
$$
(This limit corresponds to (3) with $\delta m/ m \sim 1$.)
If the mass of the heavy eigenstate satisfies this constraint,
a relativistic light state and a non-relativistic heavy
state do not separate without gravitational potential.
In halos, the states separate easily because $v_l$
is much greater than the escape velocity.

\vskip1em\vfill\break
{\bf IV. What if the mass states separate?}

The states will separate when 			\hfill\break
a) $m_h \gg m_l$ and $m_h \ge 0.16$~GeV,	\hfill\break
b) $m_h \approx m_l$ and $\delta m/m 
\ge 2.4 \times 10^{-3} m_{\rm GeV}^{-3}$.  	\hfill\break
In both cases one has to worry about the exponentially 
large number of lighter particles over the heavier ones
once the temperature drops below their mass difference.

DM particles freeze out when 
$H_f > \Gamma_f = \langle \sigma_m v_f \rangle n_f$,
where $H_f$ is the Hubble constant at freeze-out and
$\sigma_m \sim 10^{-36}$ is the cross-section of DM
interactions with ordinary matter and I estimated it
from the requirement: $\Omega_m\sim0.3$. The DM particles
remain collisional until a much later time
$H_c \sim \Gamma_c = \langle \sigma_{si} v_c \rangle n_c$,
where $\sigma_{si} \sim 10^{-24} \gg \sigma_m$.
Collisions are accompanied by conversions from
heavy states to light and back. However, if the energy
of the light component drops below the heavy particle's 
rest mass, the conversions of light to heavy are suppressed 
by the Boltzmann factor. 

One possibility to avoid this is to have the mixing
angle vanishing and, hence, no conversions will occur.
In fact, the effective mixing angle may be much smaller 
than that in vacuum due to interactions with matter 
(similar to neutrinos). But after the DM freeze-out, 
such interactions are nearly absent and this idea
does not work.

Thus the only way out is to assume that 
$\delta m \le T_c \ll T_f$, which is possible only in case (b).
(Here and below $c=\hbar=k_B=1$.) We explore this possibility below.

Assume that DM has an $s$-wave annihilation cross-section
$\langle \sigma_m v \rangle \propto T^n$ with $n=0$.
Freeze-out of cold relics occurs when
$\Gamma \sim H \simeq T^2/m_{\rm Pl}$, i.e.,
$$
\sigma_m\, v_f\, n_f \sim T_f^2/m_{\rm Pl}
\eqno{(7a)}
$$
After freeze-out, the total number of dark matter particles is 
conserved; hence their number density scales as $n\propto T^3$. 
The collisional cross-section $\sigma_{si}\sim$ constant 
and $v\propto T^{-1/2}$, i.e., $n=1/2$. 
Collisions and conversions become inefficient 
(we assume large mixing $\vartheta\sim1$) when
$\Gamma_c \sim H_c$ which now reads as
$$
\sigma_{si}\,v_f\,\left({T_c \over T_f}\right)^{1/2}\, n_f\,
\left({T_c \over T_f}\right)^3 \sim {T_c^2 \over m_{\rm Pl}}.
\eqno{(7b)}
$$ 
Dividing (7a) by (7b), we obtain
$$
T_c/T_f = \left( \sigma_m / \sigma_{si} \right)^{2/3}.
\eqno{(8)}
$$

Now, using Kolb \& Turner:
$$
x_f\approx\ln\left[0.038(n+1)
(g/g_*^{1/2})\,m_{\rm Pl}\,m\,\sigma_0\right],
$$
$$
\Omega h^2=10^9 {(n+1)x_f^{n+1}~{\rm GeV}^{-1} \over
\left(g_{*S}/g_*^{1/2}\right)\,m_{\rm Pl}\,\sigma_0}
$$
for $g\approx2$,\ $g_*\approx60$,\ $v_f\sim1$,\  
we have for the freeze-out
$$
x_f \simeq 20,~~~~~ 
T_f \simeq 5 \times 10^{-2}\, m_i~~~~~
\sigma_m \simeq \sigma_0 \simeq 5 \times 10^{-36}~{\rm cm}^2,
$$
where $m_i$ is the effective mass of the interacting flavor state.
From these expressions and equation (8), and noting that 
$\sigma_{si} \sim 2 \times 10^{11} \sigma_m$, we obtain
$$
T_c \simeq 2.4\, m_{\rm GeV}^{1/3} 
\left( \sigma_{m,-35} / \sigma_{si,-24} \right)^{2/3}
~~{\rm eV ~~~and~~~ }
x_c \sim 46.
$$
Finally, form the condition $\delta m / T_c \sim x_c$, we have
the upper limit on the mass difference:
$$
{\delta m \over m} \le 1.1 \times 10^{-7}\, m_{\rm GeV}^{-2/3}
\left( \sigma_{m,-35} / \sigma_{si,-24} \right)^{2/3}.
\eqno{(9)}
$$
If this constraint is satisfied, there is no over-production
of light states with respect to heavy. We have to compare this 
condition with (b) in the beginning of this section, that is
$$
0.6 \times 10^{-3} m_{\rm GeV}^{-3}\sigma_{si,-24}^{-1} 
\le 1.1 \times 10^{-7}\, m_{\rm GeV}^{-2/3}
\left( \sigma_{m,-35} / \sigma_{si,-24} \right)^{2/3}.
$$ 
This is possible for
$$
m > 40~{\rm GeV}\,
\left( \sigma_{m,-35}^2\, \sigma_{si,-24} \right)^{1/7},
{\rm  ~~~which~yields~~~ } 
\delta m/m \le 4 \times 10^{-9}.
$$
Previously, we have estimated (from $v\sim v_{\rm escape}$) the mass 
difference, which is required to significantly alter the structure 
of a dark halo, see equation (4). Both conditions for $\delta m/m$ 
are satisfied simultaneously for $v\sim10^{-4}$ 
which corresponds to about 40 km/s. Clearly this is smaller than 
the escape or even circular velocity, but, it seems, that it is not 
too small to rule out this case with confidence. Moreover, some 
changes in the cross-sections may affect this constraint. 
We conclude that this last scenario is somethat less probable 
than the scenatios in section III, and numerical N-body simulations 
are required to confirm this result.

\vskip1em\vfill
{\bf V. Conclusions}

It was shown that the light particle over-production may be
avoided in two cases. 				

{\bf (1)} Flavor states do not separate into mass states in time. 
This system behaves similar to the conventional SIDM. This scenario 
requires the DM mass to be $\le 17$~GeV in the extreme degenerate 
case [constraints (3) and (4) must be satisfied as well] and to be 
$\le 0.16$~GeV in the non-degenerate case.	

{\bf (2)} In the opposite case, mass states separate from each other
and the only plausible way out is the highly degenarate case.
It requires DM mass $\ge 90$~GeV and results in rather small 
particle velocities (less than the escape velocity). Therefore
the halo will not loose mass in this scenario, however, 
particle conversions may ``puff up'' the halo and hence
change its inner structure. 

N-body similations are needed to study the structure formation 
and evolution of halos in both cases, but especially for (2)
to, possibly, rule it out.

\end{document}